\documentclass[nonacm]{acmart}
\usepackage{pgf}
\usepackage{scalefnt}
\usepackage{tablefootnote}

\usepackage{booktabs}   
\usepackage{subcaption} 

\usepackage{algpseudocode}
\usepackage{listings}
\usepackage[ruled,vlined]{algorithm2e}
\usepackage{tabularx}
\usepackage{amsthm}
\usepackage{subcaption}
\usepackage{caption}
\theoremstyle{definition}
\newtheorem{definition}{Definition}[section]
\setcopyright{none}
\copyrightyear{2021}
\acmYear{2021}
\acmDOI{}

\acmJournal{JACM}
\acmVolume{xx}
\acmNumber{xx}
\acmArticle{xx}
\acmMonth{8}
\makeatletter
\def\blfootnote{\xdef\@thefnmark{}\@footnotetext}
\makeatother
\begin{document}

\title[]{
Joint Program and Layout Transformations to enable Convolutional Operators on Specialized Hardware based on Constraint Programming
}         




\author{Dennis Rieber}
\email{DennisSebastian.Rieber@de.bosch.com}        
\orcid{nnnn-nnnn-nnnn-nnnn}             
\affiliation{
	\position{}
	\department{}              
	\institution{Corporate Research, Robert Bosch GmbH}            
	\streetaddress{}
	\city{}
	\state{}
	\postcode{}
	\country{Germany}                    
}

\author{Axel Acosta}
\email{Axel.Acosta@de.bosch.com}
\orcid{nnnn-nnnn-nnnn-nnnn}             
\affiliation{
	\position{}
	\department{}              
	\institution{Corporate Research, Robert Bosch GmbH}            
	\streetaddress{}
	\city{}
	\state{}
	\postcode{}
	\country{Germany}                    
}
         
\author{Holger Fr\"oning}
\orcid{0000-0001-9562-0680}             
\affiliation{
  \position{}
  \department{Institute of Computer Engineering}             
  \institution{Heidelberg University}           
  \streetaddress{ }
  \city{}
  \state{}
  \postcode{}
  \country{Germany}                   
}
\email{holger.froening@ziti.uni-heidelberg.de}         


\begin{abstract}
The success of Deep Artificial Neural Networks (DNNs) in many domains created a rich body of research concerned with hardware accelerators for compute-intensive DNN operators. However, implementing such operators efficiently with complex hardware intrinsics such as matrix multiply is a task not yet automated gracefully. Solving this task often requires joint program and data layout transformations. First solutions to this problem have been proposed, such as TVM, UNIT or ISAMIR, which work on a loop-level representation of operators and specify data layout and possible program transformations before the embedding into the operator is performed. This top-down approach creates a tension between exploration range and search space complexity, especially when also exploring data layout transformations such as im2col, channel packing or padding.

In this work, we propose a new approach to this problem. We created a bottom-up method that allows the joint transformation of both compuation and data layout based on the found embedding. By formulating the embedding as a constraint satisfaction problem over the scalar dataflow, every possible embedding solution is contained in the search space. Adding additional constraints and optmization targets to the solver generates the subset of preferable solutions.

An  evaluation using the VTA hardware accelerator with the Baidu DeepBench inference benchmark shows that our approach can automatically generate code competitive to reference implementations. Further, we show that dynamically determining the data layout based on intrinsic and workload is beneficial for hardware utilization and performance. In cases where the reference implementation has low hardware utilization due to its fixed deployment strategy, we achieve a geomean speedup of up to $\times2.813$, while individual operators can improve as much as $\times170$.

\end{abstract}


\maketitle
\section{Introduction}

Deep Learning has established itself as a pervasive method, including domains like image recognition, speech and natural
language processing, robotics, and is continuously extending further. 
Deep Artificial Neural Network (DNNs) are currently dominant and convolutions form their computationally intensive core operator, together with activation and pooling operations. However, it cannot be foreseen that this trend will continue. Instead, the community continues to innovate,
introducing extensions or novel concepts to improve accuracy and generalization of machine learning methods. Dilation or depth-wise convolutions reduce computational load, while transformers or capsules are completely new DNN operators.

Most tools in this context, such as PyTorch~\cite{Paszke2017AutomaticDI} or TensorFlow~\cite{Abadi2016TensorFlowAS}, focus on the network operator level, eg. convolutions or activation functions, with highly optimized library implementations of the individual layers. This creates a tight coupling between the DNN architecture and the targeted hardware. The authors of a recent publication~\cite{Barham:2019:MLS:3317550.3321441} identify this coupled design philosophy as a pain point in DNN research. It can lead to suboptimal training and inference runtime while researching new types of layers or hardware accelerators for which no optimized implementation is available. In turn, this will render extensive explorations regarding applicability, generalization and accuracy infeasible.

When targeting general-purpose hardware, such as GPUs and CPUs, automatic tuning tools like AutoTVM~\cite{AutoTVM:10.5555/3327144.3327258}, Ansor~\cite{Ansor:258858}, Telamon~\cite{Telamon:hal-01655602} or Chameleon~\cite{Chameleon:48873} can help finding well-performing implementations automatically.
However, resource-constrained
settings prompt the need for specialized hardware
due to the computational demand of most DNNs. 
For specialized hardware, among others, tooling is required to automate the
mapping step from abstract problem descriptions, e.g. using computational graphs, to the ISA of the underlying hardware intrinsic. This step is non-trivial as it requires program and data layout transformation to match hardware instrinsics with complex dataflows and hundreds or thousands of parallel and sequential operations over multidimensional input and output arrays. 

To improve the programming of specialized hardware, tools like TVM~\cite{DBLP:journals/corr/abs-1802-04799}, ISAMIR~\cite{DBLP:journals/corr/abs-1810-09958}, UNIT\cite{weng2021unit} or LIFT~\cite{Steuwer:2017:LFD:3049832.3049841, Mogers-Lift-Mapping} offer semi- or fully-automated approaches to embed instructions at the loop level. They are enabled by structured program transformations. Abstracting at loop level is attractive, since it allows a very concise representation of DNN workloads with billions of individual operations. Embedding instructions at this level requires pattern matching of loops and access functions in the workload against an instruction. If an embedding is not possible, program transformations create different implementation candidates, each performing an equivalent computation. Then the embedding is attempted on the new candidates. The complexity of this top-down approach was studied by Rieber and Fr\"oning~\cite{Rieber:2020:ITEM}, motivating research on novel methods for embeddings. 
One drawback is the loop-level abstraction itself, which introduces implicit implementation decisions like loop structure or access function notation into the embedding process.

Matching program structure and hardware instrinsic alone is often not enough. The data layout feeding the hardware instric is also crucial. A simple example is a matrix multiplication with a transposed operand that should be implemented with a GEMM instruction. The matching algorithm needs to either detect the transposed operand based on the access function and perform the transposition after the embedding, or perform a transpose as part of a search strategy and then attempt the matching. The matching based on TVM's abstract syntex tree (AST), for example, would fail without a prior transpose.

Transformation selection and ordering introduce additional challenges to the embedding problem, since it is not always possible to directly detect a specific sequence of transformations that leads to an embedding. Ultimately, this can lead to a non-deterministic search process where many different implementation candidates need to be generated. Additional transformations increase the complexity further, with diminishing returns on the number solutions.
Strategies to handle this complexity include performing transformations in a static order, restricting the set of possible transformations or even specifying a full template for each operator. Relying on explicit program rewrite strategies also hinders the exploration of possible solutions, since a whole subclass of implementations could be hidden behind an unavailable transformation.

This work presents a bottom-up approach based on Data-Flow Graphs (DFG), that by design contains all possible embeddings of an instrinsic into a workload.
From this exhaustive space, the subspace of legal solutions for a specific intrinsic is described using constraints. 
After finding the desired embedding, the program and data layout for the targeted accelerator can be generated together. Specifically, we present:
\begin{itemize}
	\item A method of describing the search space of the embedding problem as a Constraint Satisfaction Problem (CSP) on the level of individual scalar operations, instead of loops. This removes many implicit implementation decisions from the representation, such as loop and data layout. 
	\item A method to describe and search for desired solutions in this search space, using Constraint Programming (CP). This allows control over the solution space without changing the underlying problem formulation. This is paired with a  rule-based method to connect data-flow embeddings to joint program and data layout transformation.
	\item An evaluation of our method using VTA, a programmable architecture template that is designed for Deep Learning (DL) applications and 
	programmed with TVM. We demonstrate that our approach can generate implementations competitive to the TVM reference and how a dynamic data layout affects the overall performance.
	\item The automatic generation of novel 2D convolution implementations for different versions of the VTA accelerator, with trade-offs among operator performance, memory footprint and tensor transformation efficiency.
\end{itemize}

In this work we focus on the generating code and data layout for the buffers feeding the accelerator's data paths. Handling multi-level memory hierarchies is considered beyond the scope of this work. Further, it is currently restricted to workloads and hardware ISA with perfect loop-nests and no conditions in the control flow. However, since we use the polyhedral model as the underlying representation, support for more complex workloads and intrinsics is possible by principle. While the evaluation of this work focusses on the inference of convolutional workloads, an extensions to MLPs and similar workloads is feasible. Workloads in a training scenario are subject for future research.
In section, \ref{sec:related} a general overview of DNN deployment tools is presented. Section \ref{sec:method} introduces the general methodology and program representation of our proposed approach, while section \ref{sec:embedding-cp} explains how CP serves as a flexible and extensible tool to solve the embedding problem. We evaluate our approach on VTA in section \ref{sec:strict}. We demonstrate how the bottom-up method can recreate the results of an existing reference implementation, while offering additional flexibility during code generation. Finally, section \ref{sec:relaxed} shows how rule based strategy generation can offer implementations outperforming the reference on metrics like memory footprint or operator performance.

\section{Related Work} \label{sec:related}
Existing DL frameworks such as TensorFlow~\cite{Abadi2016TensorFlowAS} or Pytorch~\cite{Paszke2017AutomaticDI} focus on application-level interfaces to describe DNNs with a set of operators, including convolutions, pooling and activations. These operators are mapped to libraries like cuDNN~\cite{Chetlur2014cuDNNEP} for NVIDIA GPUs or NNPack \footnote{\url{https://github.com/Maratyszcza/NNPACK}, accessed 05.2021} for x86 architectures. These libraries offer handcrafted kernels with high performance for a specific hardware target.
Intel nGraph~\cite{DBLP:journals/corr/abs-1801-08058} bridges to hardware using transformers, containing all hardware-specific optimizations. In the case of x86, it uses libraries. 
The specificity of these approaches provides near-optimal performance on specific hardware, but increases the engineering effort when exploring new operators or hardware architectures.
\begin{table}[t]
	\centering
	\caption{Comparing a selection of DNN compilation tools for hardware acceleration.}
		\label{tab:tools}
\begin{tabularx}{\textwidth}{l l l c c l}
	 &  &  & \textbf{Data Layout} & & \\ 
	\textbf{Name} & \textbf{Workloads}  & \textbf{IR} & \textbf{Transform} & \textbf{Embedding} & \textbf{Optimization} \\ 
	 \hline
	TVM~\cite{DBLP:journals/corr/abs-1802-04799}& CNN, MLP, LSTM & Loops & $\times$ & $\times$ & AutoTVM~\cite{AutoTVM:10.5555/3327144.3327258} \\ 
	DORY~\cite{Dory} & CNN, MLP & Loops + Caches & $\times$ & $\times$ & CP \\ 
	 
	ISAMIR~\cite{DBLP:journals/corr/abs-1810-09958}& CNN, MLP, LSTM & Loops + Registers & transpose & \checkmark & static \\ 
	 
	UNIT~\cite{weng2021unit}& CNN, MLP & Loops + Registers & $\times$ & \checkmark & AutoTVM \\ 
	Lift~\cite{Steuwer:2017:LFD:3049832.3049841,Mogers-Lift-Mapping}& CNN, MLP & Functional Lang.  & $\times$ & \checkmark & auto tuning \\ 
	This Work & CNN & Polyhedral + DFG & rule-based & \checkmark & CP + AutoTVM \\ 
\end{tabularx} 
\end{table}

Further discussion of related work is split into three groups. The first group targets DNN accelerator with a fixed dataflow ISA. We consider this group of tools be the closest to our work.
To add context, the second group explores data and operation mapping in dataflow and CGRA architectures and the third group discusses dynamic FPGA configurations for DNNs. While all groups are concerned with DNN acceleration on spezialed hardware, the optmization goals and challenges are different, which is why we consider a direct comparison not useful.
\paragraph{Tools targeting ISA accelerators}
Table \ref{tab:tools} presents a selection of tools used to bring CNNs to hardware accelerators with an ISA interface, for example TensorCores in NVIDIA GPUs ~\footnote{\url{https://www.nvidia.com/en-us/data-center/tensor-cores}, accessed 05.2021} or the VTA~\cite{VTA} hardware.  \textbf{Data Layout Transform} describes if and what changes the tool can make to the data layout and \textbf{Embedding} marks if the tool can automatically rewrite the program as necessary to place the hardware intrinsic in the workload. Lastly, \textbf{Optimization} describes how the implementation is optimized before deployment.

TVM~\cite{DBLP:journals/corr/abs-1802-04799} is an open-source compiler stack for DNNs, describing networks in a functional language called  Relay~\cite{Roesch2018RelayAN}. Individual operators for execution are lowered to a scheduling language inspired by Halide~\cite{Ragan-Kelley:2017:HDA:3176926.3150211}. Feedback-based performance optimization, or auto tuning, is then used to find good schedules for individual operators~\cite{AutoTVM:10.5555/3327144.3327258}. Schedules determine parameters like loop tiling or vectorization for each loop.
Custom hardware backends are programmed by embedding hardware intrinsics into the AST. This \textit{embedding} is only semi-automatic. For every operator an expert has to specify an embedding \textit{strategy} which statically defines the data layout  and binds the instructions loops instruction to workload loops. Based on this, code for individual operators can be generated and optimized. Static strategies are limited in their ability to adjust to different operator layouts or parametrizations and are not part of AutoTVM's optimization space. Especially if a dimension in the instruction is larger than the dimension in this specific operator instance, workarounds like zero-padding are necessary, but reduce hardware utilization.

DORY~\cite{Dory} is on the same level auf embedding automation as TVM, as both data layout and the calls to accelerating functions are done by hand. Instead of auto tuning, a constraint programming based optimization process searches for the best cache hierarchy utilization in the PULP~\cite{pulp} hardware.

ISAMIR~\cite{DBLP:journals/corr/abs-1810-09958} automates the embedding problem at loop level by matching tensor dimensions and artihmetic operations in the loop nest to the ISA, striving to derive loop orderings and data layout from the embedding attempts. If this is not possible, a non-determistic program transformation is performed. The resulting program is then matched against the ISA, again. How the tool deals with too-small dimensions on for the embedding is not reported, among others.

Similar to ISAMIR, UNIT ~\cite{weng2021unit} automates the embedding process based on arithmetic operations and register adressing. It verifies that the addressed data between mapped dimensions between workloads and instruction strictly match each other. Program rewrites are limited to embedding and optmization. If no possible embedding is present in the workload, the deployment will fail. For example, in order to embed NVIDIA's Tensor Core intrinsic the workload needs be specified in the im2col format by a human expert, before an embedding can be successful.

Rewrite systems creating different implementations for same computation are also used by LIFT~\cite{Steuwer:2017:LFD:3049832.3049841}, for scheduling as well as exploring possible embedding of specialized hardware instrinsics into DNN operators~\cite{Mogers-Lift-Mapping}.

Except for ISAMIR no discussed tool is concerned with automatic data layout transformations. ISAMIR itself only reports the ability to transpose operands, if necessary. Determining the layout before the embedding can lead to cases where no embedding is possible (see UNIT). Having the layout as part of a top-down embedding search space can increase the complexity of the problem exponentially ~\cite{Rieber:2020:ITEM}.
This work offers an alternative embedding strategy by embedding the instruction on the dataflow level and then derive how the data layout and program need to be transformed together. We support data layout transformation like transpose, padding or dimension packing, but also enable the dynamic fusion of multiple tensor dimensions into a single dimension for the hardware instrincs to map against. Features in the CSP results can potentially be matched to arbitraty data layout rewrites. This way, complex data layout rewrites like im2col or parallelizing a sliding window are possible.

Like all tools in this group, we split the problem into strategy choice (data layout and algorithm) and schedule optimization for multiple reasons. i) ~Treating the problem as a combined optmization problem would increase the complexity exponentially. ii)~Runtime performance of individual operators is often not only the goal when selecting the strategy. Memory footprint, latency or data movement are often conflicting goals during deployment. iii) ~In the context of network-level optimizations ~\cite{TASO:10.1145/3341301.3359630}, strategies need to be coordinated between adjacent operations.

\paragraph{Tools targeting dataflow accelerators}
The second group is about tools targeting dataflow architectures. This class of accelerators enables dynamic data routing in the hardware. Determinig the temporal and spatial data movement through the accelerator is the main objective of this group. Since there is no ISA, an embedding process is not necessary. Further, global data layout transformations are not explored, as they are mainly concerned with data locality in the processing elements, not outside of it.  
Timeloop~\cite{Timeloop:8695666} has a focus on on loop-level program rewrites for dataflow based accelerators such as Eyeriss~\cite{Eyeriss:10.1109/ISCA.2016.40} and DianNao~\cite{DianNao:conf/asplos/ChenDSWWCT14}. Workload and hardware are abstracted over the 7 loops of a 2D convolution and user-defined annotations specify which hardware and instruction loops are possible mapping candidates. Timeloop then attempts to map the operator to the available buffer memories and processing units through successive tiling and reordering.
In a similar matter, dMazeRunner~\cite{dMazeRunner:10.1145/3358198} targets the same class of hardware and workload. Through the use of search space restrictions, they can reduce the searchtime to a few minutes. Both, Timeloop and dMazeRunner build on analytical models of energy, execution and latency to sweep the optmization space exhaustively.

While Timeloop and dMazeRunner focus on DNN workloads, Novatzki et. al. ~\cite{Nowatzki:10.1145/2491956.2462163} propose a more general approach, solving an integer linear programming (ILP) problem to find the ideal spatial placement.
Similar is the work by Chaudhuri et. al.~\cite{DBLP:conf/iccad/ChaudhuriH17}, with a SAT-based compiler for the dataflow in Coarse Grained Reconfigurable Architectures (CGRA). It uses a flow-graph abstraction and SAT solving. 
The goal is to fully compile an application with a static schedule for a CGRA hardware target. Their main contribution is the static scheduling in time and space for a CGRA with a flexible dataflow architecture.

\paragraph{Tools generating FPGA configurations}
This last group of tools dynamically generate FPGA configurations for DNN workloads. AutoSA ~\cite{AutoSA:10.1145/3431920.3439292} is a continuation of PolySA ~\cite{PolySA:10.1145/3240765.3240838}, both using the polyhedral model to compute systolic array architectures, as well the necessary spatial and temporal schedules. 
Similarly, Tabla ~\cite{TABLA:10.1145/3174243.3174257} generates FPGA accelerators for 2D and 3D DNNs based on the hardware templates for the Winograd method.
All three publication in this group use an anlytical model of the generated hardware for optimization.
\\\\
In this discussion we did not mention MLIR ~\cite{mlir} on purpose. Its philosophy of being a general point of interaction between different, more specialized tools in order to establish cooperation between different domains make it hard to classify in the terms disucssed above. Potentially, every tool introduced here could be integrated into MLIR as a distinct dialect, sharing abstractions and optimizations with other tools.

\section{Embeddings for Dataflow Graphs} \label{sec:method}
This work focusses on workloads found in the computation of DNNs, especially convolutions, operating over n-dimensionsal arrays called tensors, with bounds known at compile time. The computations performed in DNNs, such as convolutions, matrix multiplications or pooling, consist of deep loop nests without conditional statements and thus are highly structured. This allows the usage of concise notations for computations, like the matrix multiplication $A[i,j] = \sum_k X[i,k] \cdotp Y[k,j]$ as a \textit{tensor expression}. These expressions can be directly translated into a loop-based program.
However, existing work~\cite{Rieber:2020:ITEM} also showed that using loop-level abstractions to embed instructions into operators requires explicit transformations of the program before an embedding can be attempted. This creates a large search space to find a correct sequence of program transformation that result in an embeddable version of the program.
Here, we propose a bottom-up approach to the problem: by analyzing how instructions and operators fit to each other on a scalar level, the necessary transformation for an embedding can be inferred automatically. This removes the need for top-down, non-deterministic decision making, as used in previous approaches.
\subsection{Dataflow Graphs}\label{subsec:dfg}
Conceptually, our approach is based on dataflow graphs (DFG). A DFG represents every scalar operation necessary to perform a computation as a directed graph. Nodes represent operations, and edges the dataflow. Formally:

\theoremstyle{definition}
\begin{definition}
	A DFG is a labeled, directed graph, defined as $G = (N,E,l)$, where $N$ is the set of nodes, the set of directed edges is $E \subseteq  NxN$ and $l()$ is a function assigning labels from  the set $L_N \cup L_E$. $L_N = \{\{Operation\}, \{Data\}\}$ is the set of node label classes, holding  tensor shapes, data types and arithmetic operations. $L_E = \{spatial,sequential\}$ is the set of edge labels.
\end{definition}
Further, a DFG has the following properties:
\begin{itemize}
	\item Nodes with only $data$ labels generate outgoing $sequential$ edges for each operation consuming the data. They have no incoming edges. They represent the input values of the computation modelled by the DFG.
	\item Nodes with $operation$ labels perform a scalar operation, consuming the data of the incoming $sequential$ edges and produce one or more outgoing $sequential$ edges. 
	\item Commutative reduction operations are modelled by a $sequential$ self-edge. This optimization can reduce the number of nodes and edges in a graph significantly, without loosing correctness of representation.
	\item Nodes with $operation$ labels  with only an outgoing self-edge, or no outgoing edges, represent the computation results.
	\item Nodes labelled $operation$ can have bidirectional $spatial$ connections to other nodes, performing the same computation, but for a different output element. These connections indicate parallelism in the computation. Potentially, $spatial$ edges lead to a set of $k$ fully connected nodes. The number of edges can be reduced by pruning the connections to create a graph with one internal node and $k-1$ leaves. In this star subgraph, the transitive property maintains the parallelism information. 
\end{itemize}
\begin{figure}[t]
	\centering
	\begin{subfigure}[b]{0.45\textwidth}

	\begin{align*}
	T: A_{i,j} = \sum_k X_{i,k} \cdot Y_{k,j}
	\end{align*}

	\caption{Tensor Expression for a matrix multiplication}
	\label{fig:te}
	\end{subfigure}
	\hfill
	\begin{subfigure}[b]{0.45\textwidth}

	\begin{lstlisting}[frame=single, language=python, numbers=left, numberstyle=\tiny,tabsize=2,commentstyle=\color{gray},mathescape]
for i in I:
  for j in J:
    for k in K:
      tmp = X[i,k] * Y[k,j]
      A[i,j] += tmp
		\end{lstlisting}
	\caption{Tensor Expression (a) as a naive loop nest}
					\label{fig:te-loops}
	\end{subfigure}
	\\
	\begin{subfigure}[b]{0.45\textwidth}
		\centering
		\includegraphics[width=0.7\textwidth]{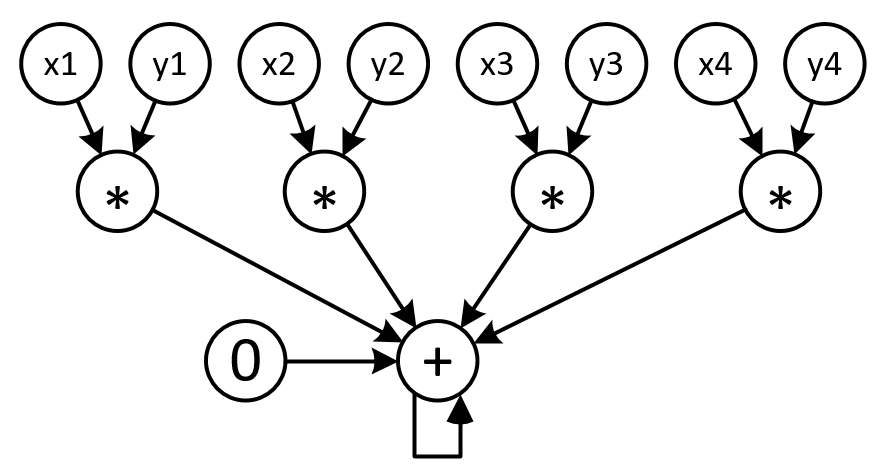}
		\caption{Partial DFG for a single output element with a contracted add operation}
		\label{fig:DFG}
	\end{subfigure}
\hfill
	\begin{subfigure}[b]{0.45\textwidth}
	\centering
	\includegraphics[width=0.5\textwidth]{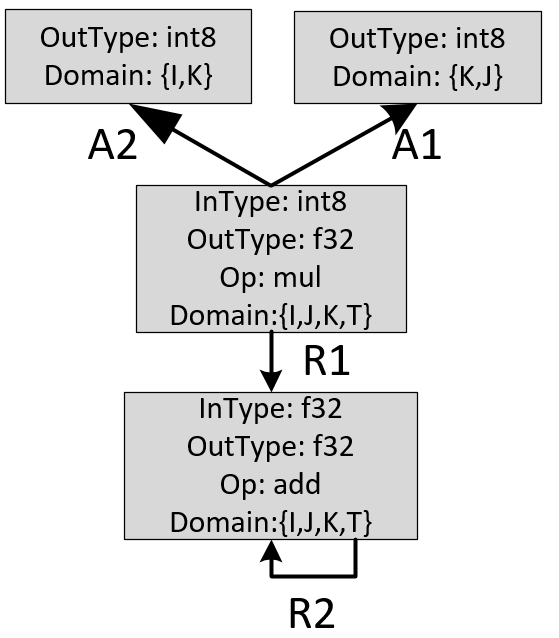}
	\caption{Tensor Expression graph in set and relation based representation}
	\label{fig:expr-graph}
	\end{subfigure}
	\caption{Matrix multiplication expression, its dataflow graph and the used representation with polyhedral sets and relations}

	\label{fig:GEMM-Example}
\end{figure}

For the tensor-based workloads in a DNN, every output element is computed by the same sequence of operations, but with a different subset of input values.
Figure \ref{fig:DFG} shows the partial dataflow graph of a 4x4 matrix multiplication. Each subgraph computing one output element shares a set of input nodes with its neighbours but no intermediate results. This property removes the need to cover the full operator DFG with a sequence of instructions, but instead solves the problem for a small subset and then uses a hardware-dependent inference step to determine the structure of the full program. We will demonstrate this in sections \ref{sec:strict} and \ref{sec:relaxed}.

Embedding an instruction DFG $G_i = (N_i, E_i, l_i)$ into an operator DFG $G_o = (N_o, E_o, l_o)$ is equivalent to the subgraph isomorphism problem. For an embedding we need to find an injective function $f:G_i \rightarrow G_o$ that describes a distinct subset of nodes and edges in $G_o$ that exactly matches $G_i$, formally described as: $\forall (s,t) \in E_i \Rightarrow (f(s), f(t))\in E_o$. This function has to maintain the labeling, such that $\forall s \in N_i: l_i(s) \equiv l_o(f(s))$.
The main advantage of this approach is that the matching problem itself is not bound to implicit implementation decisions like loop ordering or data layout of tensors or access functions. This removes the need for transformations in the search to account for these decisions. Instead, we can derive these from the result of the embedding.

However, we also identified two challenges with this approach:
\begin{enumerate}
	\item With a space complexity of $O(|N|^2)$ and $O(|N| + |E|)$, adjacency matrices and lists are not suited to represent a full DFG for operators like conv2D with billions of operations. Subsection \ref{subsec:polyhedral} explains how a polyhedral program representation is used to abstract the workload while maintaining the detailed information of the basic dataflow graph. 
	\item Solving the embedding only with subgraph isomorphism is too imprecise of a solution formulation. Often, there are additional restrictions to the available hardware and its software interface. A possible implementation needs to be inside this restricted space in order to be valid. Section \ref{sec:embedding-cp} explains how constraint programming is leveraged to overcome this problem. 
\end{enumerate}
\subsection{Program Represenation} \label{subsec:polyhedral}
The first challenge is addressed with a more concise program representation, inspired by the polyhedreal model, a powerful compiler methodology capable of expressing computations in quasi-affine loop nests. Several tools in the Deep Learning community leverage this representation to generate efficient DNNs kernels. TensorComprehensions~\cite{Vasilache2018TensorCF} targets GPU optimizations and MLIR~\cite{mlir} provides a whole polyhedral dialect for loop optimization.
Polyhedral program representation is an abstraction of loop based programs, with its components abstracting the program and data dependencies on a scalar level~\cite{523109}:
\begin{itemize}
	\item The instance set $S$ is the set of all dynamic execution instances. $S$ is described by a set of integer tuples. Each tuple $s$ describes exactly one dynamic execution instance. 
	\item The data dependence relation $D$ is the union of all binary relations between pairs of instances in $S$.
\end{itemize}
Reflecting this on our DFG representation, the node set $N_o$ of $G_o$ is $S$ and the $sequential$ edges model the data dependence relations $D$. We represent $S$ as a set of integer tuples. Every tuple is a specific operation happening in the instance set. To describe the sets of integer tuples, we use the notation
\begin{align}
\{[e_0, ..., e_n] : \tau_0, ..., \tau_n \}
	\label{form:dynamic-exe}
\end{align}
where the fixed lower and upper bounds of each tuple element $e_j$ are defined by a condition $\tau_j$. The conjunction ($\wedge$) of all $\tau$ terms contains the full set.
The data dependence relation $D$ is represented by a binary relation between two sets. A relation maps elements from the source set to the target set. Relations are denoted as 
\begin{align}
	s_{source} \rightarrow s_{target} = \{[e_0, ..., e_n] \rightarrow [e'_0, ..., e'_n] : \Phi_0, ..., \Phi_m\}
	\label{form:relation}
\end{align}
where every term $\Phi_k$ defines a condition in the relation. Elements in the target tuple are denoted $e'$. The relation condition $\Phi_k$ is used to describe any source element $e$  that maps to element $e'$ in the target set. The conjunction of all $\Phi$ terms encapsulates the whole relation domain.

To describe operators with this polyhedral representation, we move from an explicit DFG to a set-based representation. Every element from the original tensor expression $T$ (figure \ref{fig:te}), like an arithmetic operation or input tensor, is modelled by a domain set $d \subset S$, as in equation \ref{form:dynamic-exe}. For example, the domain $d_*$ contains all multiplication nodes in $G_o$. For the matrix multiply example in figure \ref{fig:te}, all dynamic execution instances and input tensors are contained in the sets
\begin{align}
	S = \{[i,j,k,t] : 0 \le i < I \wedge 0 \le j < J \wedge 0 \le k < K \wedge 0 \le t < \#T\} \\
	X = \{[i,k] : 0 \le i < I \wedge 0 \le k < K\} \\
	Y = \{[k,j]: 0 \le k < K \wedge 0 \le j < J\} 
\end{align}
where $I,J,K$ are the domain bounds in figure \ref{fig:te} and loop bounds in figure \ref{fig:te-loops}, respectively. Since the goal is to compare two different programs, we model the sequence of individual arithmetic operations explicitly. For this, $S$ contains an additional dimension $t$ that describes the order of expressions in $T$. To do this, we assign each expression (multipy, add) in $T$ an integer value. For example, we explicitly model the statements in lines 4 and 5 in figure \ref{fig:te-loops} as individual points in the dynamic instance set. Input tensors are defined by a set of their shape. Now, every node in the original DFG is contained in a union of the above sets, for the example this means $N_o \equiv S \cup X \cup Y$.

Tensor access functions and dataflow are modelled by binary relations between two domains.
There is a dataflow between two instances $(s1, s2) \in S$  if there exists a relation for which $s1 \rightarrow s2 \neq \emptyset$. 
The dataflow of the example in figure \ref{fig:expr-graph} is described by the following relations:
\begin{align}
	R1: * \rightarrow +  = \{[i,j,k,t] \rightarrow[i',j',k',t'] : i'=i \wedge j'=j'\wedge k'=k \wedge t'= t_+\}\\	
	R2: + \rightarrow +' = \{[i,j,k,t] \rightarrow [i',j',k',t'] : i'=i'\wedge j'=j \wedge k'=k+1 \wedge t'=t\}\\
	A_X: * \rightarrow Y = \{[i,j,k,t] \rightarrow [i',k'] : i' = i \wedge k' = k \wedge t=t_*\}\\
	A_Y: * \rightarrow X = \{ [i,j,k,t] \rightarrow [k',j'] : j' = j \wedge k' = k \wedge t=t_*\}
\end{align}

Relation $R1$ specifies that the multiplication and addition happen in the same loop iteration, but are ordered by their textual position in $T$, specifically that the node following the multiplication needs to be an add operation. $R2$ can be interpreted such as two add operations happen sequentially in iteration dimension $k$ and that the same add operation is performed. This is the self-edge in the original DFG. To accommodate commutative operations, the term controlling the reduction order can be relaxed. The $spatial$ edges of the original DFG are modelled in the same way. $A_X$ and $A_Y$ encode the access function itself and by which node the access is performed, in this case the multiplication in $T$. The union of all relations describes the edges of $G_o$, and specifically for the example, $E_o \equiv R1 \cup R2 \cup A_X \cup A_Y$.
Bringing the sets and relation together with the original labelling function $l_o$ creates the expression graph shown in figure \ref{fig:expr-graph}.

Not all of the relations in this representation are \textit{symmetric}, being the same in both directions. The relation $* \rightarrow X$ is \textit{surjective} and \textit{functional}, meaning that every multiplication can exactly map to the one tensor element  in $X$  it consumes. However, the inverse relation $X \rightarrow *$ is \textit{non-functional}. Every input element in $X$ is used by multiple multiplications, but not all of them. The relation of $A_X$ and $A_Y$ reflects this, as they contain no term with properties for $i'$ or $j'$ respectively. As a result, for one specific input value in $X$, the relation describes the subset of all multiplications using this value.  

\section{Embedding as a Constraint Satisfaction Problem}\label{sec:embedding-cp}

Based on the concise yet detailed program representation from section \ref{sec:method}, we will now discuss how constraint programming (CP) solves the second challenge presented in section \ref{subsec:dfg}. CP is a method of solving constraint satisfaction problems (CSPs). By describing the space of legal embeddings in terms of constraints, a solver can automatically generate solutions belonging to this space, if any exists. Adding more constraints makes the solution space more specific, while relaxing constraints can serve as a tool for implementation strategy exploration.
The choice of constraint programming is motivated by its expressiveness in the program formulation and customizable propagation and search algorithms.

\theoremstyle{definition}
\begin{definition} \label{def:csp}
A CSP is formally defined as triple $\langle X,D,C \rangle$, where
\begin{itemize}
	\item $X = \{x_j | 0 \le j \le n\}$ is a set of variables, for which we have to find a value.
	\item $D = \{d_j | 0 \le j \le n\}$ is the set of value domains, from which we assign values to the respective variables.  An assignment $Asn(d_j, x_j): x_j = v$ selects value $v \in d_j$ for $x_j$ to take. Variable $x_j$ can only ever receive values from domain $d_j$, not from any other domain in $D$.
	\item $C = \{c_i | 0 \le i \le m\}$ is the set of constraints. A constraint $c_i$ is formed over a subset of variables $g_x \subset X$ and evaluates if all assignments $Asn(g_d, g_x)$ with $g_d \subset D$ are valid. 
\end{itemize}
The CSP is satisfied when all assignments are performed and no assignments 
violate the conjunction of $C$.
\end{definition}

Once the problem is modelled with variables and constraints, a solver begins the process of assigning values and evaluating constraints. Evaluating every possible assignment is infeasible in most practical applications. 
In CP, every constraint comes with a propagator removing values from the domain that cannot be part of a valid solution. The propagator is a monotonic filtering algorithm: it only removes values from a domain, but never adds any and is specific for each constraint. The propagator infers which values to remove from a domain based on the domains and assignments of other variables under the same constraint.
To find a solution, the solver uses a search algorithm to systematically perform assignments and propagate the assignments through the domains. A backtracking-based search algorithm can find all possible solutions in a given problem. 
Variable selection determines which $x \in X$ is assigned a value next. Value selection is the specific implementation of  $Asn(d,x)$. Variable and value selection impact the time-to-solution and need careful consideration when designing the constraint program.

\subsection{Problem Space Defintion}\label{subsec:Space}
This section will discuss how we represent the embedding as a CSP in $GeCode$\footnote{https://github.com/Gecode/gecode}, the solver used for this work.
\theoremstyle{definition}
\begin{definition} \label{def:embedding-space}
The full space of the embedding problem is:
\begin{itemize}
	\item $X = \{x | \forall x \in N_{i}\}$ For every node of the instruction DFG a variable is created. Therefore, every scalar operation and data element in the instruction is represented by a variable.
	\item The set of domains is defined as $D = \{d |d \subseteq S_{operator} \}$. Every domain is a subset of the operators instance set in the polyhedral representation. The subset is determined by the node type. The domains of nodes labelled $data$ are the shape of the respective tensors. The domain of nodes labelled $operation$ is the instance set.
\end{itemize}
This formulation describes the embedding on the scalar level. Since every potential assignment between nodes in the instruction and nodes in the workload is described by this formulation, it also contains every possible solution to the embedding problem.  
\end{definition}

Over the space defined in \ref{def:embedding-space} we can formulate constraints that specify solution properties.
The most important constraint is matching the dataflow between operator and instruction. With other constraints we can further reduce the solution space. 

\subsection{Embedding Constraints}\label{subsec:constraints}
This section presents the constraints placed upon the problem space defined in the previous section.
\subsubsection{Subgraph Isomorphism} Solving this problem with CP is a well-researched problem~\cite{hal-01381438,zampelli2007,larrosa_valiente_2002} and solutions range from direct formulations to highly optimized implementations. For this work we directly model the instruction DFG $G_i$ we want to discover in the operator's target graph $G_o$, as shown in figure \ref{lst:post-subgraph}. As described in definition \ref{def:embedding-space}, every node of $G_i$ is a variable. Every edge $(s,t) \in E_i$ is then modelled with a binary constraint describing the dataflow (line 3). For better propagation, we also model the spatial edges (line 5). To fully express isomorphism we use a global \textit{AllDiff} constraint such that every node can only occur once in the solution (line 7). 
\begin{figure}
\centering
\begin{subfigure}[b]{0.47\textwidth}
\begin{lstlisting}[frame=single, language=python, numbers=left, numberstyle=\tiny,tabsize=2,commentstyle=\color{gray},mathescape]
for (s,t) in $E_i$:
 if label(s,t) $\in$ sequential: 
   edge(s,t, etype $\Leftarrow$ sequential)
 else:
   edge(s,t, etype $\Leftarrow$ spatial)
  
AllDiff($N_i$)
\end{lstlisting}
\caption{Describing the instruction graph as pairwise $edge$ constraints.}
\label{lst:post-subgraph}
\end{subfigure}
\hfill
\begin{subfigure}[b]{0.47\textwidth}
\begin{lstlisting}[frame=single,language=python, numbers=left, numberstyle=\tiny,tabsize=2,commentstyle=\color{gray},mathescape]
if l(s) $\ne$ l(s.val)
  return Failed
# evaluate data dependence
rel $\Leftarrow$ eval(l(s)$\rightarrow$l(t), s.val) 
if rel = $\emptyset$: 
  return Failed
t.domain $\Leftarrow$ t.domain $\cap$ rel

if t.size = 1:
  t.val $\Leftarrow$ t[0] # assign solution
  return Finished
return # value of t selected by solver
\end{lstlisting}
\caption{Propagation of edge constraints using the data dependence relation}
\label{lst:propagate-edge}
\end{subfigure}
\label{lst:props}
\caption{Algorithms for describing subgraph isomorphism and computing the propagation based on the polyhedral data dependence relations}
\end{figure}

Now that the problem is described, the solving process can begin. During this, the solver eventually assigns a variable a value from its domain. Assigning a value means to select one node of $N_o$ as a possible candidate to match a node in $N_i$. The propagation algorithm in figure \ref{lst:propagate-edge} then checks the label of the variable against the node assigned from the domain (lines 1-2) and if possible also filters the other node's domain (lines 4-9).
The propagator is filtering values directly based on the data dependence relations in $T$.
It evaluates the relation (line 4) and removes values from the partner node's domain where no connection exists (line 7). If the relation between the pair is \textit{functional}, it can directly assign a solution (line 9-11). 
Even if this is not the case, the propagation is powerful enough to \textit{subsume} the domain, meaning that only valid solutions for this constraint remain in the domains and no further propagation is necessary. The remaining domain values are evaluated with respect to the other constraints over their variable. If evaluating the relations leads to an empty domain, the assignment fails (line 6). 
Finally, when values are assigned to both $s$ and $t$, the constraint checks for correctness by verifying there is an edge in the $G_o$ connecting the pair, or formally $Asn((s, t),(d_s, d_t)) \in E_o$.

To further aid the propagation we also model the implied parallel edges in the dataflow graph (line 5 in \ref{lst:post-subgraph}). Since fully expressing this would result in a large number of constraints, we leverage the transitive property of the pairwise constraints. We pick an arbitrary node in the DFG and add a constraint to every parallel node it has. If now any node parallel to the first node gets assigned a value, the domain of first node is pruned to only contain nodes parallel to the assignee. This, in turn, propagates to all other nodes parallel to the first node. While this introduces a degree of indirection in the propagation, the number of pruned values remains the same.

\subsubsection{Axis-parallel hyper-rectangle constraint} 
\begin{figure}[t]
	\begin{lstlisting}[frame=single, language=python, numbers=left, numberstyle=\tiny,tabsize=2,commentstyle=\color{gray},mathescape]
$m_k$ $\Leftarrow v_1 - v_0$
if $\frac{m_k}{|m_k|} \notin V_B$:	return Failed
	
$V_B \Leftarrow V_B \cap \frac{m_k}{|m_k|}$	
$V \Leftarrow V \cap v_0$	
$DimTable  \Leftarrow \emptyset$ 		
$LiveTable[m_k] \Leftarrow 1 $
for $v_n \in V$:
	$move \Leftarrow v_n - v_{n-1}$
	if $move \in LiveTable$:
		LiveTable[move] =+ 1
		if not VerifyAndReset(LiveTable, DimTable): return Failed
	# check if this is a dimension jump
	else if $\frac{v_n - v_0}{|v_n - v_0|} \in V_B \wedge move = (v_n - v_0) + (v_0 - v_{n-1})$:
		DimTable[$m_k$] $\Leftarrow$ LiveTable[$m_k$] # size of $d_{k-1}$
		LiveTable[move] $\Leftarrow$ 0 # Add counter for new outermost dimension $d_k$
		$m_k$ $\Leftarrow$ move # remember diagonal move of $d_k$
		$V_B \Leftarrow V_B \cap \frac{v_n - v_0}{|v_n - v_0|}$		
	else: return Failed 
DimTable[$m_k$] $\Leftarrow$ LiveTable[$m_k$]
return BoundingBox(DimTable)	
	\end{lstlisting}
	\caption{Algorithm for hyper rectangle inference. $V$ is the list of variables, each variable holding a point and $V_B$ is the vector base of the domain's shape (e.g. shape of the input tensor).}
	\label{lst:hrec}
\end{figure}

For this work, but also for DNN workloads and accelerators in general, regular memory access patterns are a sensible restriction on the solution space. Especially convolutions operate in high-dimensional, rectangular spaces, or hyper-rectangles. Selecting an axis-parallel subset of this domain enables common data layout transformations, like transposing, fusing or tiling. At the same time, this helps reducing the size of the solution space.
This constraint is developed to match and propagate the shape of a rectangle with any number of dimensions $n > 0$ from an ordered tuple of points $V = [v_0, ..., v_n]$. After only a few decisions the propagator can infer a bounding box from the selected points and the total number of points in $V$. By intersecting this bounding box with the domain, we efficiently remove values that can never lie within the rectangle. The constraint supports regular strides in any dimension.  
For example, after selecting only two points along one axis of the tensor, we can bound this dimension to $ bound = \#V \cdot |v_0 - v_1|$. 

After the initial step determining the innermost dimension (line 1-7), the algorithm in fig. \ref{lst:hrec} performs a linear iteration of $V$, trying to infer if the points create a rectangle with an arbitrary number of dimensions. 
If the points describe a rectangle in lexicographic order, the vectors from one point to next can be split into two classes. A \textit{step} is the movement from one element in the innermost dimension to the next. The other type of movement is a \textit{jump}, where the iteration jumps into the next line of the innermost dimension, moving diagonally through the rectangle. For every dimension in the mapping, the \textit{step} and \textit{jump} are identical and happen a fixed number of times.  The algorithm iterates all points, increasing a counter for every known \textit{step} or \textit{jump} (lines 10-14). After a counter reaches the size of its dimension, it rolls back to zero. The verification (line 12) checks if for a \textit{jump} into dimension $d_k$, all counters of the inner dimensions $d_{k-1}...d_0$ are zero. If one counter is non-zero, the \textit{jump} breaks the regular structure of the hyper-rectangle.
Every \textit{jump} $\notin LiveTable$ possibly adds a new dimension to the rectangle (line 14). To maintain rectangle properties, the \textit{jump} vector  $v_n - v_{n-1}$ has to be as the same as $(v_n - v_0) + (v_0 - v_{n-1})$,  where the normalized $(v_n - v_0)$ has to be one of the tensor's base vectors $V_B$. The restriction to the elements of $V_B$ ensures right angles at the corners and that the rectangle is axis aligned. Every dimension of the operator tensor can be used exactly once, which is enforced by removing the normalized vector from $V_B$. After a new \textit{jump}, the new outermost dimension $d_k$ is added to shape of the rectangle. Now we also know that the size of dimension $d_{k-1}$ is the value of its counter (lines 14-18). The length of each rectangle side is stored in $DimTable$.
After iterating all points, the $DimTable$ is used to compute the bounding box (line 21). Since this constraint is called during the solving process, it is possible that not all values of $V$ are assigned. In this case, formula (10) estimates the bound of dimension where the size is not yet known.
For brevity, we left out sanity checks for the correct size of dimensions, for example in line 15 the size of a dimension has to be an even divisor of the points in the rectangle: $LiveTable[d_{k-1}] \bmod \#V = 0$. 

\begin{figure}[t]
	\centering
	\includegraphics[width=\textwidth]{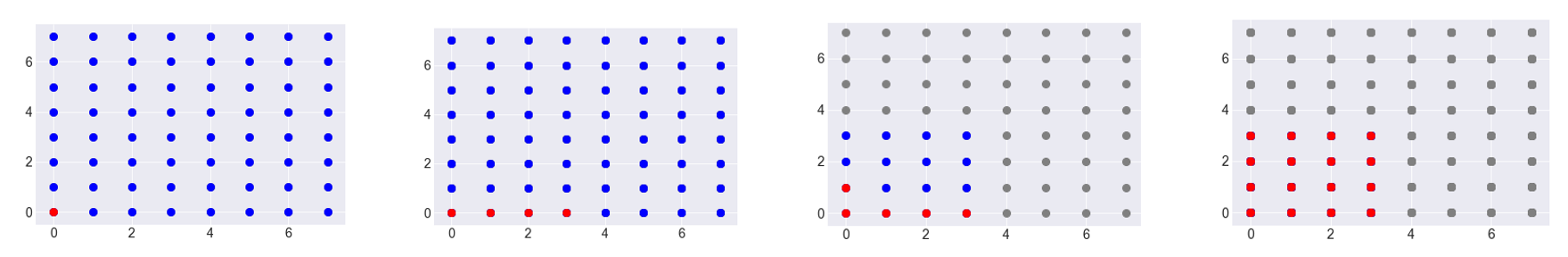}
	\caption{From left to right the plots above show how propagation and assignment happens in the rectangle constraint. Blue dots represent domain values, red ones assigned variables, and grey ones the values removed by propagation.}
	\label{fig:hrec}
\end{figure}

Figure \ref{fig:hrec} provides an example for the inference and resulting propagation. The blue points represent the domain, the red points the selection for one of 16 variables. For the first 4 steps (first and second from left) there is no option to propagate, since the size of the dimension along the $x$ axis (8) is smaller than the total number of variables (16). However, after the \textit{jump} we can start removing values for $x$ and $y$. In this example, the removed values are grey.
We can remove every value $(y,x)$ where $x > 3$, since this is the size of our innermost dimension $d_x$. From the value $(1,0)$ selected for the fifth variable, the propagator can infer that the expansion into the $y$ dimension can be no larger than:
\begin{align}
	d_y = \frac{\#V }{\prod_{i=0}^{k-1} d_i \cdot stride_i}
\end{align}
In this case this would be $d_y =\frac{\#V}{d_x \cdot 1} = \frac{16}{4  \cdot 1} = 4$.
Since the algorithm operates on a set of points, this process is agnostic to the dimension ordering in workload and tensor, making the mapping rotation invariant. This is allows us to project a found mapping into the memory shape necessary for code generation.

\subsubsection{Memory Access Functions} This constraint also affects which input values can be part of the solution. However, it is much simpler than the previous constraint. It specifies which memory access patterns are allowed. For example it is possible to forbid access patterns like stencils to be part of the solution, or access patterns with regular strides and offsets. 
For this work we implemented a simple check for linear memory access. It computes if inputs assigned from the operators are all in dimensions with a single iterator and a constant stride. However, the polyhedral model allows for much more powerful memory analysis, if necessary.

\subsubsection{Padding}
To support a wider range of workloads zero-padding can be added to individual dimensions of the iteration domain. This happens when either a) an individual dimension is below a specified threshold or b) the dimension is not an even divisor of dimensions in the hardware intrinsic. This gives the CSP solver more freedom in terms of the search and possible candidates. By penalizing padding in the later candidate selection, solutions with unnecessary padding are discarded early.

\subsection{Branching Strategies}
The previous sections described how the problem is modelled in CP, now we will discuss the search used in the actual solving process.
Variable and value selection strategies determine how the problem space is explored by determining which variable is assigned a value next. To better fit the underlying problem the strategies can be customized. To reduce the amount of choices necessary during branching it is desirable to trigger as much propagation as possible with every assignment, such that every domain gets subsumed as early as possible.

We use a variable selection strategy based on groups of node types in $G_i$. From the example in figure \ref{fig:expr-graph}, all multiplications are in group $g_*$. Changing the order of groups can result in varying degrees of propagation. When starting with a group of input nodes, less propagation is possible due to the non-functional relations to their consuming nodes. Starting with $g_*$, every node assigned a value automatically propagates to its inputs as well as to the following add operation.
Our implementation currently begins with the output variables and propagates backwards through the DFG, which proved to be a robust heuristic for short solver runtimes. The value selection strategy implements a lexicographic search through the domain.

\subsection{Candidate Selection}
Padding and other data transformations necessary for an embedding can create an overhead in the number of multiply-accumulate operations (MAC) and data movement. 
We add an optimization term to the constraint problem, minimizing the MAC and data movement overhead $O_{MAC}$ and $O_{Data}$, compared to the theoretical minimum:
\begin{align*}
	O_{MAC} = MAC_{Total}  - MAC_{min} \hspace{4em}
	O_{Data} = Data\ movement_{Total} - Data\ movement_{min}
\end{align*}
In our work, we used the number of tensor elements for the data movement computation, not the absolute size.
When treating $O_{MAC}$ and $O_{Data}$ as elements of a vector $\vec{\textbf{o}} = \lbrack O_{MAC},O_{Data}\rbrack^T$,
the CSP optimizes the term $min(||\vec{\textbf{o}} \cdot \vec{\textbf{w}}||)$, where $\vec{\textbf{w}}$ is a user-defined weight vector. It can be used shift the impact of the overheads on the candidate selection, depending on the on the used hardware and technology. This eliminates solutions with padding or other overheads, when possible. 
\section{Validation Experiments}\label{sec:strict}
In this section we validate our approach against TVM on a series of convolutional workloads. To allow for direct comparisons, we only allow embeddings and program rewrites equivalent to ones used by TVM. This is also in line with the reported abilities of  UNIT~\cite{weng2021unit}, LIFT~\cite{Steuwer:2017:LFD:3049832.3049841, Mogers-Lift-Mapping} and ISAMIR~\cite{DBLP:journals/corr/abs-1810-09958}. Numerical correctness in all experiments is ensured by comparing the results of our approach with the reference implementations. Since all operands and results are quantized to $int8$, we observed no issues with floating-point associativity.

\subsection{Experimental Setup}
The evaluation is performed on VTA~\cite{VTA}, a hardware accelerator providing a matrix-multiply (GEMM) instruction with a corresponding processing unit, as well a vector unit for activation and scaling tasks. The hardware is instantiated on a Zynq Ultrascale+ FPGA. We use the default configuration provided by the authors with a 256kbyte weight buffer, a 128kbyte data buffer and a GEMM core with $8bit$ multiplications and $32bit$ accumulation. The GEMM unit computes $C_{xy} += A_{xz} \cdot B_{zy}^T$ with $(x,y,z) = (1,16,16)$ and its result can be processed by a vector-scalar unit for activation and quantization operations. Notice that matrix operand $B$ is transposed. The hardware has a load/store direct-memory access (DMA) unit for independent memory accesses of matrix operands. It can read and write full 2D operand matrices stored consecutively in memory.

Evaluation workloads are from the Baidu DeepBench Inference Benchmark Suite\footnote{\url{https://github.com/baidu-research/DeepBench}, accessed 01.2021}, providing 108 convolution operators from a range of domains, like image and speech processing. Twelve convolutions cannot be executed on the VTA accelerator without additional zero padding. Section \ref{sec:relaxed} will discuss possible solutions for this problem in detail.
Furthermore, 28 layers in the convolution benchmark cannot be processed by \textit{GeCode}, the solver we implemented our approach in. It uses the default C++ $int$ data type representation of the used compiler, which is 32 bits in our case. Large convolutions can yield domains substantially larger than this limit. Additional 6 layers caused various errors in the TVM compiler, like a VTA instruction buffer overflow. This leaves 62 layers to benchmark.

The conv2d reference embedding of TVM maps the three axes $x,y,z$ of the GEMM unit statically to the batch $n = x$, output channel $oc = y$ and input channel  $ic = z$ dimensions of the convolution. Each of these dimensions is split and moved to be the innermost dimensions of the \textit{input}, \textit{weight} and \textit{out} tensors.
This process is specified in a static template, applying the necessary transformations during code generation.
\begin{figure}[t]
	\centering
	\includegraphics[height=0.25\textheight]{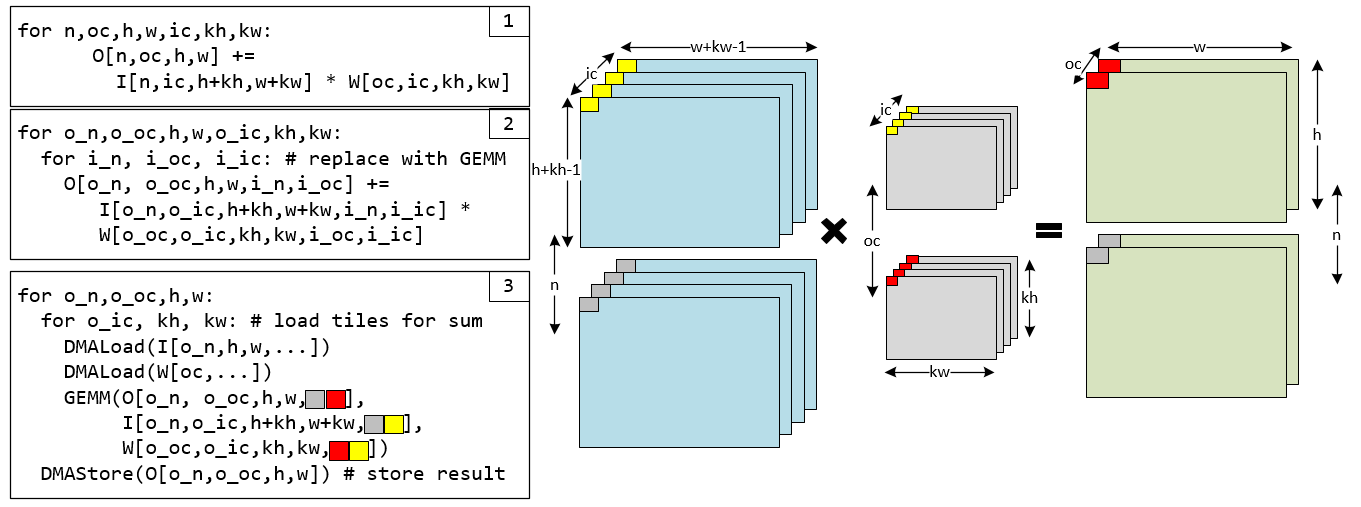}
	\caption{2D Convolution and the reference GEMM implementation}
	\label{fig:gemm-ex}
\end{figure}
Figure \ref{fig:gemm-ex} shows the loop and data layout transformations for a convolution with a 2D memory tiling, where each tile is a matrix operand that can be loaded/stored by the DMA. Step 1 is the baseline conv2d. The loops and tensor are tiled as explained before (fig. \ref{fig:gemm-ex}, step 2). The resulting implementation reorders $n_i,oc_i,ic_i$ to be the innermost loops. These loops are then replaced by a call to the GEMM instruction.
The DMA load and store operations are then placed around the operation to continuously load values until an output segment is complete (Figure \ref{fig:gemm-ex} step 3). The TVM convolution implementation expects a NCHW data layout.
Implementing other workloads follows a similar pattern -- input and output dimensions are tiled into matrices and moved to the inside. 

When TVM encounters an operator that can be accelerated by the hardware, the implementation is handled by the specified deployment strategy. A strategy implements the operator optimized for the target in TVM's IR. We build on TVM's code generation for VTA by integrating our approach into VTA's deployment strategy.
It generates the instruction DFG $G_i$ based on the hardware configuration. From there it formulates the constraint program as explained in section \ref{sec:embedding-cp}. The nodes of $G_i$ become the variables, the operator's dynamic instance sets the domain. We use the following constraints to generate mappings similar to the reference :
\begin{itemize}
	\item \textbf{subgraph isomorphism}: Match the dataflow of the GEMM instruction. This is the central constraint of the embedding problem.
	\item \textbf{hyper-rectangle}: Ensure all input and output elements are mapped into an axis aligned shape. This allows simpler memory transformations based on transpose and reshape operations.
	\item \textbf{allDiff}: Prevent the same dynamic execution instance from appearing multiple times in the same instruction call.
	\item \textbf{fixed origin}: The first match of all input and output tensors is fixed to the origin of the respective domain.
	\item \textbf{dense}: No input or output tensor is allowed to have a stride in any dimension.
	\item \textbf{linear memory access}: Only allow matches in workload dimensions with a linear memory access. This excludes, for example, strides and stencil patterns.
\end{itemize}
We can use this constraint program to attempt to embed the VTA instruction into any workload, not just convolutions. 
The solver produces a list of tuples, describing how each operation and data element in $G_i$ maps to a node in $G_o$. 
The regularity of DNN workloads like convolutions allows us to extrapolate an implementation from this information. In the solution, the variables associated with the input and output values are evaluated to compute which dimension of the instruction is matched to which dimension in the workload and what the tiling factors are. For VTA, the code generation is then straight forward. The matched dimensions are tiled by the determined factors and moved to be the innermost dimensions and loops are reordered in the same fashion. These transformations are necessary to embed the GEMM instruction. Loops and tensor dimensions not part of the embedding are free to be transformed for further performance optimization. These optimizations include loop tiling, ordering or fusion. AutoTVM is used to automatically determine the best optimization parameters for every conv2d layer. Finally, code with embedded instructions is generated by TVM's VTA programming tool flow. 

\begin{figure}[t]
	\centering
	\resizebox {\textwidth} {!} {
		\input{img/strict_nchw.pgf}
	}
	\caption{Speed-up as ratio of time of the TVM reference to time of this work, for the conv2d layers of the Baidu DeepBench Inference Benchmark. The grey envelope and dashed blue lines are the normalized standard deviation over the reference and this work, respectively. Results show that the baseline of this work based on strict mapping and standard NCHW data layout is of comparable performance to the TVM reference, with all but two layers being inside one standard deviation of the reference.}
	\label{fig:nchw-latency}
	\resizebox {\textwidth} {!} {
		\input{img/strict_nhwc.pgf}
	}
	\caption{Speed-up as ratio of time of TVM reference (NCHW) to time for a generated NHWC data layout, for the conv2d layers of the Baidu DeepBench Inference Benchmark. Except for four layers, results demonstrate consistent performance advantages and thus indicate the potential of flexible code generation for DNN operators.
	}
	\label{fig:nhwc-latency-rel}
\end{figure}
We validate our approach by comparing the performance of a micro-benchmark with the TVM reference implementation. The benchmark performs tensor packing, convolution, activation and tensor unpacking. It is implemented as a Relay~\cite{Roesch2018RelayAN} program. Packing and unpacking are performed by the ARM host CPU on the Zynq board. For the convolution operator, the TVM reference uses an expert-made implementation template that specifies which memory and loop transformations are necessary, as shown in Figure \ref{fig:gemm-ex}. This template expects a NCHW tensor layout. Our method automatically generates TVM code based on the found embedding and a specified target memory layout. As intended by the constraint program design, the implementation found by our solvers maps the instruction dimensions to the same workload dimensions as the reference. From this, a Relay program for the tensor packing is generated automatically. Both the reference and our generated solutions use AutoTVM to optimize performance. To ensure comparability, both versions use the same tiling configuration found by AutoTVM.

Our tool achieved performance competitive with the TVM reference, as demonstrated in Figure \ref{fig:nchw-latency}. After a warm-up, we averaged over 200 measurements for each layer. 
Across the benchmark, all but two layers perform within the reference's standard deviation ($\sigma$) envelope. The  mean $\sigma$  is $26ms$ for the reference as well as our solution. The absolute differences between the two approaches range from $0.1ms$ to $90ms$. These results show that our approach can compete with existing, expert-made implementations for a hardware accelerator target.

\subsection{Dynamic Data Layout}
Dynamically changing the tensor layout of a DNN for global performance optimization, such as changing from NCHW to NHWC, for instance, is a topic of interest and already supported some by existing tools~\cite{Ansor:258858}\cite{234946}. However, for accelerators like VTA, only some of the dimensions are free to be rearranged, as the packed, innermost dimensions are necessary for the embedding.
Since the solver determines which dimensions are necessary, and thus which ones are free, the memory layout of the free dimensions can be changed during code generation.

We demonstrate this by generating code for the NHWC format, including code for tensor packing and unpacking.
Figure \ref{fig:nhwc-latency-rel} compares the results of this experiment to the TVM reference, based on NCHW. Over all layers of the tested micro benchmark, the reference outperforms NHWC in only four cases. This effect correlates with the ratio between channels and image size,  as layers with larger channels show better performance in the NHWC layout. This can be explained with the data movement during the layout transformation. The packing moves data from the $ic$ and $oc$ to be the innermost tensor dimension. When dimensions are closer to their target position, the read access has less stride, yielding better CPU cache utilization.

\section{Case Studies on Dynamic Strategy Generation} \label{sec:relaxed}
In the previous section we demonstrated that our approach can achieve performance competitive to the existing reference, without having to specify a data layout manually.  This section explores scenarios where static strategies can become inefficient due to mismatches between hardware instrisic and deployment strategy and how dynamic rewrite strategies based on embedding results help improve latency and data movement.
The first scenarios explores workload variations like dilated or low-channel convolutions. The second seconario is concerned with processing element changes and how they influence the deployment. In all experiments we used a weight vector of $\vec{\textbf{w}} = \lbrack 1,1 \rbrack^T$ in the candidate selection.

Relay functions implement all required rewrites prior to the workload. The function unrolling stencils uses `relay.take()`, a gather operation copying values in an index list, as no direct im2col operator is available in Relay. All rewrites are specified in table \ref{tab:rewrite-rules-extended}. 
\begin{table}[t]
	\caption{Data layout rewrites for dynamic convolution strategy generation. The order and implementation of rewrites is manually defined for specific hardware targets and workloads. Stencil unroll and image pack are mutually exclusive per mapped dimension, which is ensured by the CSP. The hardware intrinsic dimension is $d$, workloads dimensions are $n,i$ and $k$. $s$ is the extend of $d$ in $n$ or $i$. The notation $d_e$ is the access to element $e$ in $d$.}
	\label{tab:rewrite-rules-extended}
	\begin{tabularx}{\textwidth}{c l l l l} 
		Order & Strategy & Feature & Data Layout & Access Function \\
		\hline
		1 & Stencil Unroll &  $d_e \cap d_{e+1} \ne \emptyset \wedge stride = 1$ & $ \lbrack N,I \rbrack \Rightarrow \lbrack N,I,K \rbrack $ & $ \lbrack n,i+k \rbrack \Rightarrow \lbrack n,i,k \rbrack $ \\
		1 & Image Pack &  $d_e \cap d_{e+1} = \emptyset \wedge stride > 1$ & $ \lbrack N,I \rbrack \Rightarrow \lbrack N*s,\frac{I}{s} \rbrack $ & $ \lbrack n,i+k \rbrack \Rightarrow \lbrack n,i+k \rbrack$  \\
		2 & Pad &  $d = i | i \% split \ne 0$ & $\lbrack N,I \rbrack \Rightarrow \lbrack N,I+pad \rbrack $ & $ \lbrack n,i \rbrack \Rightarrow \lbrack n,i \rbrack$ \\
		3 & Split & $d = i | i \% split = 0$ & $\lbrack N,I \rbrack \Rightarrow \lbrack N,\frac{I}{s},s \rbrack $ & $\lbrack n, i \rbrack \Rightarrow \lbrack n,i_{out},i_{in}  \rbrack$ \\
		4 &	Reorder & <hardware specific> &  $\lbrack N,I \rbrack \Rightarrow \lbrack I,N \rbrack $ & $\lbrack n, i \rbrack \Rightarrow \lbrack i,n  \rbrack$  \\
		5 & Fuse & $d = (i,j)$ & $\lbrack N,I \rbrack \Rightarrow \lbrack N \cdot I \rbrack $ & $\lbrack n,i \rbrack \Rightarrow \lbrack f_{ni}  \rbrack$ \\
	    \hline
	\end{tabularx}
\end{table}
\subsection{Low-Channel and Dilated Convolutions}
In the previous sections, twelve convolution layers in the benchmark could not be executed on the VTA without zero padding the $ic$ dimension. We will refer to them as "low-channel" convolutions. Depth-wise convolutions pose a similar problem to VTA, as they also lack the necessary number of input channels. To this set of workloads we added two dilated convolutions.

As explained in section \ref{sec:strict}, the $ic$ dimension is split with a factor the size of dimension $z$ in the instruction. If $ic < z$, padding $ic$ with $ic - z$ additional elements is necessary to generate code. However, padding results in lower utilization and larger tensors. Dilated convolutions are processed as by packing the original image and process it as independet dense covolution before interleaving the data back into the original shape. As a reference, we use a stencil that is blown up with zeros to match the dilated access pattern.
For the evaluation, we selected the five best implementations based on the selection metric in section \ref{sec:embedding-cp} as candidates for auto tuning.
\begin{table}[t]
	\caption{Generated implementations with the best \textit{combined} performance, i.e. time for transformation and operator. All numbers are reported relative to the TVM reference with padding.}
	\label{tab:relaxed-total}
	\begin{tabularx}{\textwidth}{l c c c c c c c c } 
		&& Op.  & Transf. & \multicolumn{2}{c}{Combined }  & \multicolumn{3}{c}{Memory}  \\\cmidrule(lr){5-6}\cmidrule(lr){7-9}
		Data, Weight, Pad, Stride&Dilation & S  & S & S & $\sigma$   &  Data & Weights & Tot.  \\
		\hline
(1, 700, 161, 1)(32, 1, 20, 5)0,2&1&$\times$117.697&$\times$0.095&$\times$48.089&17.998&$\times$2.330&$\times$0.100&$\times$2.268 \\
(2, 700, 161, 1)(32, 1, 20, 5)0,2&1&$\times$104.199&$\times$0.070&$\times$35.411&14.042&$\times$2.330&$\times$0.100&$\times$2.299 \\
(4, 700, 161, 1)(32, 1, 20, 5)0,2&1&$\times$170.802&$\times$0.020&$\times$27.686&10.109&$\times$0.974&$\times$0.100&$\times$0.968 \\
(1, 480, 48, 1)(16, 1, 3, 3)1,1&1&$\times$1.139&$\times$0.038&$\times$0.272&0.080&$\times$0.977&$\times$0.111&$\times$0.972 \\
(1, 108, 108, 3)(64, 3, 3, 3)1,2&1&$\times$1.485&$\times$0.053&$\times$0.618&0.153&$\times$1.000&$\times$0.444&$\times$0.974 \\
(1, 224, 224, 3)(64, 3, 3, 3)1,1&1&$\times$1.023&$\times$0.037&$\times$0.466&0.126&$\times$1.004&$\times$0.444&$\times$0.998 \\
(2, 224, 224, 3)(64, 3, 3, 3)1,1&1&$\times$1.193&$\times$0.013&$\times$0.244&0.058&$\times$3.964&$\times$0.444&$\times$3.944 \\
(1, 224, 224, 3)(64, 3, 7, 7)3,2&1&$\times$10.793&$\times$0.053&$\times$1.137&0.301&$\times$0.513&$\times$0.327&$\times$0.502 \\
(2, 224, 224, 3)(64, 3, 7, 7)3,2&1&$\times$3.310&$\times$0.046&$\times$0.876&0.213&$\times$0.513&$\times$0.327&$\times$0.508 \\
(1, 151, 40, 1)(32, 1, 20, 5)8,2&1&$\times$13.599&$\times$0.463&$\times$6.973&2.474&$\times$0.786&$\times$0.100&$\times$0.549 \\
(1, 700, 161, 1)(64, 1, 5, 5)1,2&1&$\times$11.338&$\times$0.089&$\times$3.380&1.046&$\times$0.963&$\times$0.160&$\times$0.952 \\
(2, 700, 161, 1)(64, 1, 5, 5)1,2&1&$\times$11.355&$\times$0.066&$\times$2.737&0.904&$\times$0.963&$\times$0.160&$\times$0.958 \\
(1, 18, 18, 304)(448, 304, 3, 3)0,1&2&$\times$2.550&$\times$0.139&$\times$1.829&-&$\times$1.0&$\times$0.360&$\times$0.378 \\
(1, 72, 72, 208)(304, 208, 3, 3)0,1&4&$\times$23.770&$\times$0.486&$\times$18.369&-&$\times$1.0&$\times$0.074&$\times$0.189 \\
\hline
\hline
Geo Mean&&$\times$8.788&$\times$0.069&$\times$2.813&&$\times$1.121&$\times$0.188&$\times$0.0882 \\
	\hline
	\end{tabularx}
\footnotesize{$^a$ Operator, $^b$ Transformation, $^c$ Speed-up, $^d$ Total}\\
\end{table}
\begin{table}[t]
	\caption{Generated implementations with the smallest \textit{memory footprint}. All numbers are reported relative to the TVM reference with padding.}
	\label{tab:relaxed-memory}
	\begin{tabularx}{\textwidth}{l c c c c c c c c } 
		&& Op.  & Transf. & Combined  & \multicolumn{4}{c}{Memory} \\ \cmidrule(lr){6-9}
		Data, Weight, Pad, Stride& Dilation & S  & S & S  &  Data & Weights & Tot. & $\sigma$\\
		\hline
(1, 700, 161, 1)(32, 1, 20, 5)0,2&1&$\times$116.952&$\times$0.049&$\times$30.692&$\times$0.974&$\times$0.160&$\times$0.952&0.650 \\
(2, 700, 161, 1)(32, 1, 20, 5)0,2&1&$\times$103.393&$\times$0.047&$\times$26.665&$\times$0.974&$\times$0.160&$\times$0.963&0.655 \\
(4, 700, 161, 1)(32, 1, 20, 5)0,2&1&$\times$170.802&$\times$0.020&$\times$27.686&$\times$0.974&$\times$0.100&$\times$0.968&0.622 \\
(1, 480, 48, 1)(16, 1, 3, 3)1,1&1&$\times$1.139&$\times$0.038&$\times$0.272&$\times$0.977&$\times$0.111&$\times$0.972&0.694 \\
(1, 108, 108, 3)(64, 3, 3, 3)1,2&1&$\times$1.398&$\times$0.046&$\times$0.558&$\times$0.509&$\times$0.444&$\times$0.506&0.774 \\
(1, 224, 224, 3)(64, 3, 3, 3)1,1&1&$\times$1.023&$\times$0.037&$\times$0.466&$\times$1.004&$\times$0.444&$\times$0.998&1.061 \\
(2, 224, 224, 3)(64, 3, 3, 3)1,1&1&$\times$0.201&$\times$0.037&$\times$0.168&$\times$1.004&$\times$0.444&$\times$1.001&1.181 \\
(1, 224, 224, 3)(64, 3, 7, 7)3,2&1&$\times$10.793&$\times$0.053&$\times$1.137&$\times$0.513&$\times$0.327&$\times$0.502&1.138 \\
(2, 224, 224, 3)(64, 3, 7, 7)3,2&1&$\times$3.310&$\times$0.046&$\times$0.876&$\times$0.513&$\times$0.327&$\times$0.508&1.098 \\
(1, 151, 40, 1)(32, 1, 20, 5)8,2&1&$\times$3.041&$\times$0.242&$\times$2.191&$\times$0.352&$\times$0.160&$\times$0.286&1.145 \\
(1, 700, 161, 1)(64, 1, 5, 5)1,2&1&$\times$11.329&$\times$0.023&$\times$1.112&$\times$0.963&$\times$0.160&$\times$0.952&1.127 \\
(2, 700, 161, 1)(64, 1, 5, 5)1,2&1&$\times$1.816&$\times$0.015&$\times$0.569&$\times$0.963&$\times$0.160&$\times$0.958&1.105 \\
(1, 18, 18, 304)(448, 304, 3, 3)0,1&2&$\times$2.550&$\times$0.139&$\times$1.829&$\times$1.0&$\times$0.360&$\times$0.378&-\\
(1, 72, 72, 208)(304, 208, 3, 3)0,1&4&$\times$23.770&$\times$0.486&$\times$18.369&$\times$1.0&$\times$0.074&$\times$0.189&-\\
\hline
\hline
Geo Mean&&$\times$6.068&$\times$0.053&$\times$1.938&$\times$0.765&$\times$0.209&$\times$0.643& \\
		\hline
	\end{tabularx}
\end{table}

Tables \ref{tab:relaxed-total} and \ref{tab:relaxed-memory} show the performance of solutions found by our solver regarding  overall time for operator and transformation and the memory footprint. All is reported relative to a naive padding strategy. Memory transformations and operator speed-ups are reported individually and combined. We report no variance for dilated workloads, since only a single implementation was found, respectively.
This overview shows that is possible to improve memory footprint and overall performance, but often not at the same time. Therefore, optimizing for another objective often leads to a different implementation. A more detailed view reveals that the trade-offs between the implementations we generated versus simple padding are complex and need detailed consideration for individual cases.

One of the main drivers of better inference performance is an effective hardware utilization, controlled by the padding. The largest speed-ups are achieved in layers with $ic = 1$. For $ic < z $, only $\frac{ic}{z} \cdot (h \cdot w)$ elements in the input image meaningfully contribute to the result. This drives down the effective utilization of the hardware and gives our approach the advantage.

However, padding in $ic$ and the resulting change in the number of operations alone does not give the full picture when discussing performance. Computing the number of operations in a convolution is the product of all its loops. Since the reference only pads the $ic$ dimension, the maximal possible factor increasing the number of operations is 16 with our VTA instance. This does not explain the extreme speed-ups in the layers 0-2. Increasing $ic$ does not only increase the size of the data tensor, but also generates larger weight tensors. In layers 0-2 and 7-11 the padding creates weight tensors that exceed the capacity of the accelerator's on-chip weight buffer. Implementations generated with our approach mainly affect the size of the data tensor, so the accelerator can hold the full weight tensor in the on-chip buffer. 
This effect is less pronounced for the input images, since, except for layer 9, they always exceed the buffer capacity.
Ultimately, this also explains why layers with memory footprints equal or larger than the reference with padding still perform better. There is no competition for memory bandwidth capacity by weight transfers. This means data is loaded faster and results are written sooner, which in turn clears the partial result buffer quicker.

Due to the effect of the weight buffer, the data memory footprint is almost negligible for the performance in the overall system. While the stencil unrolling produces data tensors exceeding the padded tensors by factors of up to $\times2.299$ or $4.6Mbyte$, the same implementations can also provide a speed-up of up to $\times104$, or $663ms$, versus the reference. In layers 7 and 9, the best operator performance is not achieved by the implementations with the smallest data footprint. 
Comparing tables \ref{tab:relaxed-total} and \ref{tab:relaxed-memory}, we see that the final size of the weight buffer is also not directly correlated with performance of the memory transformation operation. The implementations with lowest footprint are rarely the fastest ones.

Generally, the size of the data footprint does not directly relate to the operator performance or transformation performance. We believe that the expanding of the stencils causes this. For example in layer 0 the data footprints between different solutions found by our solver differ by factor of up to $\times2.865$, or by $3.2Mbyte$, the inference performance difference is only $\times1.672$, or $2.5ms$.

The dialted convolutions perform reasonably better than the orignal with blown up stencils, even considering the relatively expense pre- and post processing of the data. Similar to the low-channel workloads, a main benefit of this method the reduction in weight buffer pressure. The weights are not artificially inflated, reducing the necessary data movement and improve the data reuse.

Our memory transformations are between $1ms$ and $60ms$ for most implementations. The simple padding + blocking of the reference is usually one to two orders of magnitude faster. There are two reasons for poor performance. First, the 'gather' index list takes up cache space and bandwidth that could be utilized for data in the stencil unrolling. Second, it does not consider cache locality effects during the linear list traversal. This makes further discussion of the memory transformation performance moot. 

\subsection{Hardware Intrinsic Variation}
\begin{table}[t]
	\centering
	\caption{Speedup of dynamically generated strategies relative to padding on an 8x8 VTA architecture.}
	\label{tab:dynamic-low-batch}
	\begin{tabularx}{.82\textwidth}{l  c c c c c c } 
		& Op.  & Transf. & Combined  & \multicolumn{3}{c}{Memory}  \\\cmidrule(lr){5-7}
	Data, Weight , Pad, Stride &  S & S & S  &  Data & Weights & Tot.  \\
		\hline
(1, 32, 8, 8) (64 32, 3, 3), 1, 1&$\times$1.60&$\times$0.89&$\times$1.13&$\times$0.60&$\times$1.00&$\times$0.77\\
(1, 32, 16, 16)(64 32, 3, 3), 1, 1&$\times$6.64&$\times$3.31&$\times$4.27&$\times$0.19&$\times$1.00&$\times$0.33\\
(1, 32, 32, 32)(64 32, 3, 3), 1, 1&$\times$12.65&$\times$6.20&$\times$7.71&$\times$0.16&$\times$1.00&$\times$0.21\\
(1, 256, 8, 8)(256, 256, 3, 3 1, 1&$\times$3.85&$\times$1.47&$\times$2.47&$\times$0.60&$\times$1.00&$\times$0.90\\
(1, 128, 16, 16)(256, 128, 3, 3), 1, 1&$\times$12.00&$\times$4.42&$\times$8.22&$\times$0.19&$\times$1.00&$\times$0.57\\
(1, 128, 32, 32)(256, 128, 3, 3), 1, 1&$\times$8.04&$\times$11.93&$\times$9.47&$\times$0.16&$\times$1.00&$\times$0.32\\
(1, 72, 56, 56)(96, 72, 1, 1), 0, 1&$\times$5.62&$\times$9.10&$\times$8.44&$\times$0.14&$\times$1.00&$\times$0.14\\
(1, 256, 7, 7)(512, 256, 1, 1), 0, 1&$\times$4.60&$\times$4.12&$\times$4.32&$\times$0.22&$\times$1.00&$\times$0.57\\
(1, 8, 224, 224)(24, 8, 3, 3), 1, 2&$\times$10.27&$\times$6.39&$\times$6.73&$\times$0.13&$\times$1.00&$\times$0.13\\
(1, 72, 56, 56)(96, 72, 3, 3), 1, 2&$\times$6.95&$\times$5.14&$\times$5.46&$\times$0.16&$\times$1.00&$\times$0.18\\
\hline
\hline
Geo Mean&$\times$6.26&$\times$4.21&$\times$4.99&$\times$0.21&$\times$1.00&$\times$0.33 \\
\hline
	\end{tabularx}
\end{table}
In this scenario we change the VTA hardware architecture from an $i,j,k = 1\times16\times16$ to a $8\times8\times8$ GEMM. This improves the data reuse and increases arithmetic intensity. As long as $i \le n$, the new hardware layout is no issue for embedding process. However, inference with batch size \textit{n~=~1} is common for many DNN use cases, so the hardware needs to support this efficiently. 
Now the static VTA template needs padding to deploy the workload, reducing the effective hardware utilization and increasing data movement. We compare this to dynamically generated stratgies from the rewrite rules in table \ref{tab:rewrite-rules-extended}. 
A more efficient strategy is decomposing the image dimensions for dense processing into independent sub-computations and fusing them into the batch. After the computation, the result is recomposed into the final image. A drawback of this method can be the need for additional padding to enable even splitting factors. A second, popular alternative strategy is im2col.
As in the previous section, we chose the top five implementations based on the selection metric in section \ref{sec:embedding-cp} as candidates for autotuning. After the optimization we selected the candidate with the best overall performance for comparison to the reference. The results are in table \ref{tab:dynamic-low-batch}. The overall improvements for an operator range from $\times1.13$ to $\times9.47$, with the operator inference alone showing improvements up to $\times12.56$. Except for the first, and smallest, workload, the data transformations are now also faster, ranging from $\times0.89$ to $\times11.93$. While our data layout changes are more expensive to perform, the increase in batch size makes the default transformations much more expensive, due to the large memory access strides. This is a similar effect as in the data layout experiment in section \ref{sec:strict}.

The theoretical improvement from $n=1$ to $n=8$ should be limited to 8, for the transformations as well as the execution. The geo-mean speedups in all three categories are within this limit. The most probable cause for invidividual workloads with faster inference is a weak auto-tuning result for the reference. For the two tranformations with a speedup $>8$, the authors suspect microarchitectural effects. The ARM core performing the operation has a total of \textit{1MByte} of L2 cache. The inputs without padding still fits into the cache, while the padded version exceeds the capacity.
But not only processing latency, also the memory footprint benefits from these measures. Through the bank we only use $\times0.21$ as much memory as the reference. The full factor of 8 is not achieved, since additional padding or data replication is sometimes necessary to accomodate the hardware, especially for workloads with smaller image sizes. There, the split in the image can't fully capture the parallelism needed for the batch, making padding necessary.

\section{Search Robustness}\label{sec:discussion}
\begin{figure}[t]

 \resizebox {1\textwidth} {!} {
	\input{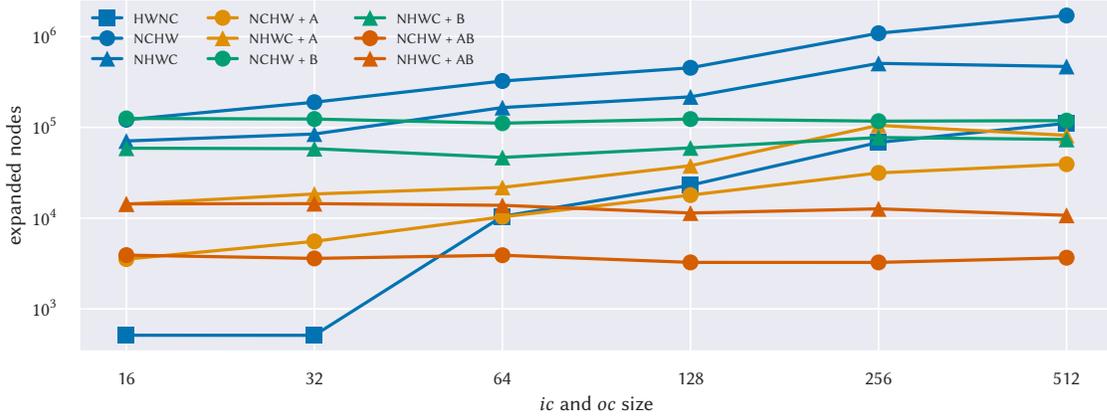}
}
	\caption{Search effort for different conv2d $ic$ and $oc$ channel sizes with various search strategies and domain layouts. A is asset search, B is domain size reduction. AB combines both strategies.}
	\label{fig:search-times}
\end{figure}
The evaluation in sections \ref{sec:strict} and \ref{sec:relaxed} showed promising results on the functionality of the proposed method. It can reproduce existing implementations, applies to new workload variations and completely new implementations can be generated by relaxing the search space. However, a weak point of this method is that the complexity scales directly with the number of operations in both instruction and workload. This section will explore and discuss this issue in conjunction with the general search robustness of constraint programming systems.

The performance of constraint programs for different problems is sensitive to the used search strategy. Changing strategies for value and variable selection can lead to exponential differences in runtime for the same problem statement. Our method amplifies this, as the variable count and domain sizes change for different problems and instructions. To demonstrate this, we explore how various operator layouts and search strategies for conv2d affect the effort for embedding a the VTA GEMM instruction with the strict solution space from section \ref{sec:strict}. 
In figure \ref{fig:search-times}, the number of search tree nodes expanded by the solver is a measure for effort to find a solution.
Without additional search strategies (A, B or AB) a large difference between different operator layouts and an upwards trend for all operator layouts is clearly visible. The ideal layout "HWNC" has a very low initial search effort. This is an artefact, as for small channel sizes (16 and 32), most initial value and variable selections in lexicographic order are correct. For larger layers, the effect dissipates and the upwards trend is the same as "NCHW" and "NHWC".
While the propagation is efficient at removing large parts of search space, not all of it can be excluded, leading to a linear search through the pruned space. The filtering is limited by the \textit{non-functional} behavior of some data dependence relations and the commutative nature of the convolution. However, this behavior is not inherently bad, as it is caused by input value reuse and commutative operations. Without these, different implementation strategies would not be possible in the first place.

We will now discuss two strategies to improve the search robustness. The first one is a straightforward approach to reduce the domain size. A group of constraint variables $g$ with the same domain, for example all variables describing the output operation, can be assigned a unary pruning constraint. This constraint thresholds the size of all dimensions in m-dimensional domain $d_m \subset S$  for all variables in $g$ by
\begin{align}
	\forall e_i \in d_m | 0 \le i  \le m : e_i = 
	\begin{cases}
		e_i & \text{if $ b_g \cdot stride \le e_i$}\\
		b_g \cdot stride  & \text{otherwise}
	\end{cases}
\end{align} where $b_g$ is the upper bound for all dimensions $e_i$ in $g$. Figure \ref{fig:search-times} reports how this method (B) stabilizes the search effort for growing domains. For this, we set $b$ to the largest dimension size in the instruction. Since $b$ removes large parts of search space, it can also remove potential solutions. Therefore, a more conservative or adaptive strategy for setting $b$ can help with exploration. Because the solver posts this constraint for every variable individually, the domain propagation happens before the solver begins the search. Therefore, this is equal to simply presenting a smaller problem to the solver. The drawback of this approach is the reduced efficiency with an increasing $b$ and $stride$, and no filtering for dimensions smaller than the threshold value. This constraint overlaps with some of the work the hyper-rectangle constraint is performing.

The second strategy changes the order in which the $n$ dimensions of the instance set $S$ are traversed. This is motivated by figure \ref{fig:search-times}, showing how different dimension orderings for conv2d affect the search. We take the operator layout (order of dimensions) directly from the workload specification and during the solving  it is traversed in lexicographic order. The layer with an ideal layout (HWNC), where the dimensions for the mapping are traversed first, arrives at a solution faster. To improve the robustness we propose an increased domain exploration diversity instead of a fixed dimension order.
A portfolio search~\cite{Portfolios:GOMES200143} uses multiple assets, where each asset has different order of searching through the $n$ dimensions in $S$. Each asset is a copy of the problem space, executed concurrently. Applying the portfolio search to the order of dimension traversal in $S$ yields a more robust search strategy. However, one asset for every possible of the $n!$ permutations of $S$ would be infeasible.
The portfolio can leverage instruction and operator properties to reduce the number of assets. The instance set $S$ is split into a number of spatial dimensions $n_s$ and reduction dimensions $n_r$. For every instruction with $k_s$ spatial and $k_r$ reduction dimensions, where $k_s < n_s$ and $k_r < n_r$, only
\begin{align}
	\#assets = \frac{n_s!}{(n_s-k_s)!} \cdot \frac{n_r!}{(n_r - k_r)!}
\end{align}
assets are necessary to create one asset with an ideal instance set layout for a lexicographic search. This strategy also helps in relaxed search scenarios, since it can potentially increase the exploration diversity.
The fact that more assets are created for instructions with more dimensions is not necessarily a drawback. Since all assets can be searched in parallel, a more complex problem could potentially be assigned more resources.

Figure \ref{fig:search-times} reports how asset-based searches (A), domain bounds (B) and their combination (AB) reduce the embedding effort. The asset-based strategy shows a clear reduction in the total effort for both operator layouts. With increasing channel size, the performance becomes comparable to a search with an ideal operator layout. Also, the absolute difference between different memory layouts is reduced. The domain bound (B) limits the effects of increasing the search effort for growing channel sizes. Its effect is especially pronounced for large channels, operators with small domains would not benefit as much from this strategy. Combining both strategies (AB) ultimately leads to a stable and fast search. The difference between data layouts in AB is solely an artefact of the assets creation and execution order.

\section{Conclusion and Future Work}
We presented a method to embed instructions offered by neural network accelerators into convolutional computations by jointly manipulating program and data layout. The approach is based on constraint programming and a polyhedral model. It solves the embedding on the scalar level, where explicit program rewrites like transpose or dimension fusion are no longer necessary to find an embedding. From the scalar embedding, program and data transformations necessary for hardware execution can be derived.  By proposing suitable variable and value selection strategies, the overall complexity of finding an embedding stays manageable, even for large DFGs with millions of nodes and edges.

Section \ref{sec:strict} showed how our method can automatically generate implementations similar to the reference. More importantly, it demonstrated how a more general approach for embedding and code generation can yield better performance when considering the necessary data layout transformations. How this translates to individual DNNs depends on the number memory packing operations necessary during execution, but the results encourage more research towards network-level optimizations for accelerators.

For operators where a static implementation template creates low hardware utilization, our approach can produce new strategies with significantly improved operator performance. Speedups of up to $\times170$ are possible. To do this, the constraints describing the solution space are relaxed and hardware-specific rules enable more complex code generation through data and program rewrites. The data in section \ref{sec:relaxed} also shows that the best implementation strategy depends on the specific operator, hardware layout and target metric. This further motivates our research on a more flexible embedding process, as finding the best implementation for every operator is non-trivial. The produced results also motivate research into the interplay of data layouts, their transformation and how this interacts with the overall DNN and accelerator architecture.

Overall, the basic principle of the approach opens many avenues for further research. For example, this work only focusses on instructions with bounded input dimensions and a fixed dataflow. Accelerators like Eyeriss~\cite{Eyeriss:10.1109/ISCA.2016.40} or DianNao~\cite{DianNao:conf/asplos/ChenDSWWCT14} offer more programming flexibility with different dataflow patterns, on-chip networks and deeper memory hierarchies.
Section \ref{sec:relaxed} showed that the general exploration of transformations that introduce a degree of inefficiency into the computation is sometimes necessary to find the best implementation. In this respect, the search for good implementations as part of the constraint solving is an open problem. Similar work has been proposed the authors of Telamon~\cite{Telamon:hal-01655602}. By pairing a constraint based search process with an analytical GPU performance model, loop tiling and other optimization parameters are determined quickly and with high result quality. However, the analytical model goes against the current trend of machine learning based optimization tools and limits applicability to different hardware architectures. The combination of a constraint model with automatically learned hardware models is a promising research direction.

\bibliographystyle{ACM-Reference-Format}
\bibliography{cp-embedding}

\end{document}